%% file: main.tex
\documentclass[twocolumn]{aastex63}

\usepackage[utf8]{inputenc}
\usepackage{url}
\usepackage{xspace}
\usepackage{natbib}  
\usepackage[frozencache,cachedir=.]{minted}
\usepackage{amsmath}

\newcommand{\pyspeckit}{\texttt{pyspeckit}\xspace}
\newcommand{\astropy}{\texttt{astropy}\xspace}
\newcommand{\hh}{\ensuremath{\mathrm{H}_2}\xspace}
\newcommand{\ammonia}{\ensuremath{\mathrm{NH}_3}\xspace}
\begin{document}

\title{{\sc{pyspeckit}:} A spectroscopic analysis and plotting package}

\author[0000-0001-6431-9633]{Adam Ginsburg}
\affiliation{Department of Astronomy, University of Florida, P.O. Box 112055, Gainesville, FL, USA}

\author[0000-0002-5327-4289]{Vlas Sokolov}
\affiliation{Max-Planck-Institut f\"ur extraterrestrische Physik, Giessenbachstrasse 1, D-85748 Garching, Germany}

\author[0000-0002-0455-9384]{Miguel de Val-Borro}
\affiliation{Planetary Science Institute, 1700 E. Fort Lowell, Suite 106, Tucson, AZ 85719, USA}

\author[0000-0002-5204-2259]{Erik Rosolowsky}
\affiliation{Department of Physics, 4-181 CCIS, University of Alberta, Edmonton, AB T6G 2E1, Canada}

\author[0000-0002-3972-1978]{Jaime E. Pineda}
\affiliation{Max-Planck-Institut f\"ur extraterrestrische Physik, Giessenbachstrasse 1, D-85748 Garching, Germany}

\author[0000-0002-3713-6337]{Brigitta M. Sip\H{o}cz}
\affiliation{IPAC, MC 100-22, Caltech, 1200E. California Blvd. Pasadena, CA 91125}

\author[0000-0001-9656-7682]{Jonathan D. Henshaw}
\affiliation{Max-Planck-Institut f\"ur Astronomie, Königstuhl 17, 69117, Heidelberg, Germany}

\begin{abstract}
\pyspeckit is a toolkit and library for spectroscopic analysis in Python. We
describe the \pyspeckit package and highlight some of its capabilities,
such as interactively fitting a model to data, akin to the historically widely-used
\texttt{splot} function in \texttt{IRAF}. \pyspeckit employs the
Levenberg-Marquardt optimization method via the \texttt{mpfit} and
\texttt{lmfit} implementations,
and important assumptions regarding error estimation are described here.
Wrappers to use \texttt{pymc} and \texttt{emcee} as optimizers are provided.
A parallelized  wrapper to fit lines in spectral cubes is included.   As part of the \astropy affiliated
package ecosystem, \pyspeckit is open source and open development and welcomes input and collaboration from
the community.

\end{abstract}

\section{Introduction and Background}
Spectroscopy is an important tool for astronomy. Spectra are represented as
the number of photons, or total energy in photons, arriving over a specified
wavelength (or equivalently, frequency or energy) range. Emission and
absorption lines caused by transitions between states in ions, atoms, and molecules bear important information
in their observed intensity, width, and velocity centroid. These parameters are
typically measured from model fits to the data, such as Gaussian, Lorentzian,
and Voigt profiles. Historically, \texttt{IRAF} \citep{iraf} provided the astronomy
community with easy-to-use tools for line fitting, but \texttt{IRAF}
development has mostly ceased in the last several years.
The lack of an equivalent available
tool in Python prompted the creation of \pyspeckit.

\pyspeckit development began in 2009 with a script called `showspec' in the
\texttt{agpy} package hosted on Google Code. It was created and used by a
graduate student to plot and sometimes fit profiles to spectra in python. At
the time, IDL was still more popular than python at most institutes
\citep[the first evidence
that python had overtaken IDL in popularity among astronomers was presented in
][]{Momcheva2015a}, and there were no publicly available and advertised
tools for spectral plotting, fitting, and general manipulation
(\texttt{astropysics}, \citealt{Tollerud2012a}, was developed contemporaneously and
solved many of the same problems as \pyspeckit). The \astropy package
\citep{AstropyCollaboration2013,AstropyCollaboration2018} had its first commit in
2011, so even the basic infrastructure for such analysis was
not yet established.

\pyspeckit's graphical user interface (GUI) features were inspired by IRAF's
\texttt{splot} tool, while the fitting features were inspired in part by \texttt{xspec}
(\url{https://heasarc.gsfc.nasa.gov/xanadu/xspec/}).  Over subsequent years,
\pyspeckit grew by incorporating more sophisticated models  and improving its
internal structure.  The package was moved out of \texttt{agpy} and into its
own repository in 2011, first spending a few years on Bitbucket in a mercurial
repository, then finally moved to GitHub in 2012, where it currently resides (\url{https://github.com/pyspeckit/pyspeckit/}).
Release v1.0 is available on Zenodo \citep{pyspeckitV1}.

Because \pyspeckit's initial development preceded \astropy, some features
were included that later became redundant with \astropy.  Most notably, \pyspeckit included
a limited system for spectroscopic unit conversion.  In 2015, this system
was completely replaced with \astropy's unit system.  Around the same time,
the Doppler conversion tools (converting from frequency or wavelength to
velocity) that existed in \pyspeckit were pushed upstream into \astropy,
highlighting the mutually beneficial role of \astropy's affiliated
packaged system \citep{AstropyCollaboration2018}.  \pyspeckit became
an \astropy affiliated package in 2017 (An affiliated package is an astronomy-related Python package that is not part of the astropy core package, and is not managed by the project but is a part of the Astropy Project community\footnote{\url{https://www.astropy.org/affiliated/}}).

In this paper we briefly outline \texttt{pyspeckit}'s architecture and highlight
its key capabilities. In Section \ref{sec:basicstructure}, we  outline the
structure of the package.  In Section \ref{sec:gui}, we describe the GUI
system.  In Sections \ref{sec:models} and \ref{sec:cubes}, we outline
\pyspeckit's cube handling capabilities and model library. 
The appendices describe parameter error estimation for Gaussians (Appendix \ref{appendix:parerrest}) and for the ammonia model \ref{appendix:parerrestammonia}).
Appendix \ref{appendix:N2Hp} describes a benchmarking test of the N$_2$H+ model against the CLASS version.
Appendix \ref{appendix:ltemodel} describes a framework for local-thermodynamic-equilibrium-based multi-transition modeling.
Finally, Appendix \ref{appendix:scouse} describes the integration of \texttt{pyspeckit} into \texttt{ScousePy}.

\section{Package structure}
\label{sec:basicstructure}
The central object in \pyspeckit is a \texttt{Spectrum}, which has
associated \texttt{data} (e.g., flux), \texttt{error} (assumed to be symmetric, 1$\sigma$, Gaussian uncertainty), and \texttt{xarr} (e.g., wavelength,
frequency, energy), the latter of
which represents the spectroscopic axis.  A \texttt{Spectrum} object has
several attributes that are themselves classes that can be called
as functions: the \texttt{plotter}, the fitter \texttt{specfit}, and the
continuum fitter \texttt{baseline}.

There are several subclasses of \texttt{Spectrum}: \texttt{Spectra}
is a collection of spectra intended to be stitched together (i.e., with
non-overlapping spectral axes, e.g., for Echelle spectra), \texttt{ObsBlock} is a collection of spectra
with identical spectral axes intended to be operated on as a group, e.g., to
take an average spectrum, and \texttt{Cube} is a 3D spatial-spatial-spectral
cube.

\subsection{Supported data formats}

\pyspeckit supports a variety of open and proprietary data formats that have
been traditionally used to store spectroscopic data products in astronomy. 
It is always possible to create a \texttt{Spectrum} object from numpy \citep{vanderWalt2011,Harris2020} arrays
representing the wavelength, flux, and error of the spectrum, but
the supported file formats listed below make the reading process easier.

\begin{itemize}
    \item ASCII: Text files with wavelength, flux, and optional error
        columns can be read using the \texttt{astropy.io.ascii} module.
    \item FITS: The Flexible Image Transport System \citep[FITS;][]{Wells1981a,Greisen2006a,Pence2010a} format is
	supported in \pyspeckit with \texttt{astropy.io.fits}.  
    FITS spectra are expected to have their spectral axis defined using the WCS
    keywords in the FITS header.  FITS binary tables with the same
    wavelength, flux, and optional error column layout as text files
    can also be read.
    \item SDFITS: Data files following the Single
	Dish FITS \citep[SDFITS;][]{Garwood2000a} convention for radio astronomy data as
	produced by the Green Bank Telescope are partly supported in \pyspeckit.
    \item HDF5: 
        If the \texttt{h5py} package is installed, \pyspeckit will support read
        access to files containing spectra in the HDF5 format, where the data
        columns can be specified using keyword arguments.
    \item CLASS: \pyspeckit is capable of reading files from some versions of
        the GILDAS Continuum and Line Analysis Single-dish Software format
        \citep[CLASS;][]{Gildas-Team2013a}.  
        The CLASS reader has been tested with data files from
        the Arizona Radio Observatory telescopes (12-m and 10-m Submillimeter
        Telescope) and the Atacama Pathfinder Experiment (APEX) radio
        telescope.
\end{itemize}

\subsection{Plotter}
The \texttt{plotter} is a basic plot tool that comprises \pyspeckit's main
graphical user interface.
It is described in more detail in the  GUI section (\S \ref{sec:gui}).


\subsection{Fitter}
\label{sec:fitters}
The fitting tool in \pyspeckit is the \texttt{Spectrum.specfit} object.
This object is a class that is created for every \texttt{Spectrum} object.
The fitter can be used with any of the models included in the model
registry, or a custom model can be created and registered.

To fit a profile to a spectrum, several optimizers are available.  Two
implementations of the Levenberg-Marquardt optimization method
\citep{Levenberg1944a,Marquardt1963a} are provided,
\texttt{mpfit}\footnote{Originally implemented by Craig Markwardt \citet{Markwardt2009}
\url{https://pages.physics.wisc.edu/~craigm/idl/fitting.html} and ported to python
by Mark Rivers and then Sergei Koposov.  The version in pyspeckit has been
updated somewhat from Koposov's version.} and
\texttt{lmfit} \citep{Newville2014}\footnote{\url{https://lmfit.github.io/lmfit-py/},
\url{https://doi.org/10.5281/zenodo.11813}}.  Wrappers of
\texttt{pymc} \citep{Salvatier2016}\footnote{\url{https://pymc-devs.github.io/pymc/}} and
\texttt{emcee}\footnote{\url{http://dfm.io/emcee/current/},
\citet{Foreman-Mackey2013a}} are also available, though these tools are better
for parameter error analysis than for optimization.

Once a fit is performed, the results of the fit are accessible through the
\texttt{parinfo} object, which is a dictionary-like structure containing
the parameter values, errors, and other metadata (e.g., information about
whether the parameter is fixed, tied to another parameter, or limited).
Other information about the fit, such as the $\chi^2$ value, are available
as attributes of the \texttt{specfit} object.

\paragraph{Optimal $\chi^2$}
Specfit computes the `optimal' $\chi^2$, which is the $\chi^2$
value computed only over the range where the model contains statistically
significant signal.  This measurement is intended to provide a more
accurate estimate of the $\chi^2$ value by excluding pixels that are
not described by the model.  By default, the function selects all pixels where
the model value is greater (in absolute value) than the corresponding error.
In principle, this optimal $\chi^2$ may be helpful for obtaining correctly
scaled errors (see Section \ref{sec:autoerror}), though this claim has never
been rigorously tested.

\subsection{Data Selection}
An important feature of the spectral fitter is the ability to select the region
of the spectrum to be fit.  This selection process can either be done manually,
using the \texttt{selectregion} method to set one or more ranges of data to
include in the fit, or interactively using the GUI.   The selected
regions are then highlighted in the plot window if one is open.

\subsection{Continuum Fitting}
The fitting process in \pyspeckit is capable of treating line and continuum
independently or jointly.  If a model includes continuum, e.g., for the case
of a four-parameter Gaussian profile that includes an additive constant, it
can be fitted through the standard \texttt{specfit} fitter.

However, it is common practice to fit the continuum independently prior to
fitting lines.  Such practice is necessary when fitting absorption lines (the equivalent width is defined relative to a normalized continuum)
and practically necessary for heterodyne radio observations where the
continuum is usually poorly measured and corrupted by instrumental effects.
Following radio convention, the \pyspeckit continuum fitting tool is called
\texttt{baseline}. This module supports polynomial, spline, and power-law
fitting.  It is common in radio astronomy to
have wide instrumental residual features in the data that need to be fitted and
removed; this process is called `baseline subtraction'.  In other wavelength
regimes, this would typically be referred to as continuum fitting or continuum
subtraction.  In practical algorithmic terms, fitting a true astrophysical
continuum and a residual instrumental baseline are indistinguishable.

\subsection{Error Treatment}
\label{sec:errtreatment}


The \texttt{Spectrum} objects used by \pyspeckit have an attached \texttt{error}
array, which is meant to hold the $1\sigma$ independent Gaussian errors on
each pixel.  While this error representation may be a dramatic
oversimplification of the true errors for almost all instruments (since it
ignores correlations between pixels), it is also the most
commonly used assumption in astronomical applications.

The \texttt{error} array is used to determine the best-fit parameters and their uncertainties (see
\S \ref{sec:fitters}).  They can be displayed as error bars on individual
pixels or as shaded regions around those pixels using different display modes.

A typical example is given below, where we generate a spectrum and error
array using \texttt{numpy} \citep{harris2020array} and \texttt{astropy} tools \citep{AstropyCollaboration2013,AstropyCollaboration2018}.

\vspace{2mm}
\begin{minipage}{\linewidth}
\begin{minted}{python}
from astropy import units as u
import numpy as np
import pyspeckit

xaxis = np.linspace(-25, 25)*u.km/u.s
sigma_width = 3.0*u.km/u.s
data = 5*np.exp(-xaxis**2 /
                (2*sigma_width**2))*u.Jy
error = np.ones_like(data) * 0.2
sp = pyspeckit.Spectrum(xarr=xaxis,
                        data=data,
                        error=error)

sp.plotter(errstyle='fill')

sp.plotter.savefig("example_fig_1.pdf")
\end{minted}
\end{minipage}

\begin{figure}[!htp]
\includegraphics[scale=1,width=3.25in]{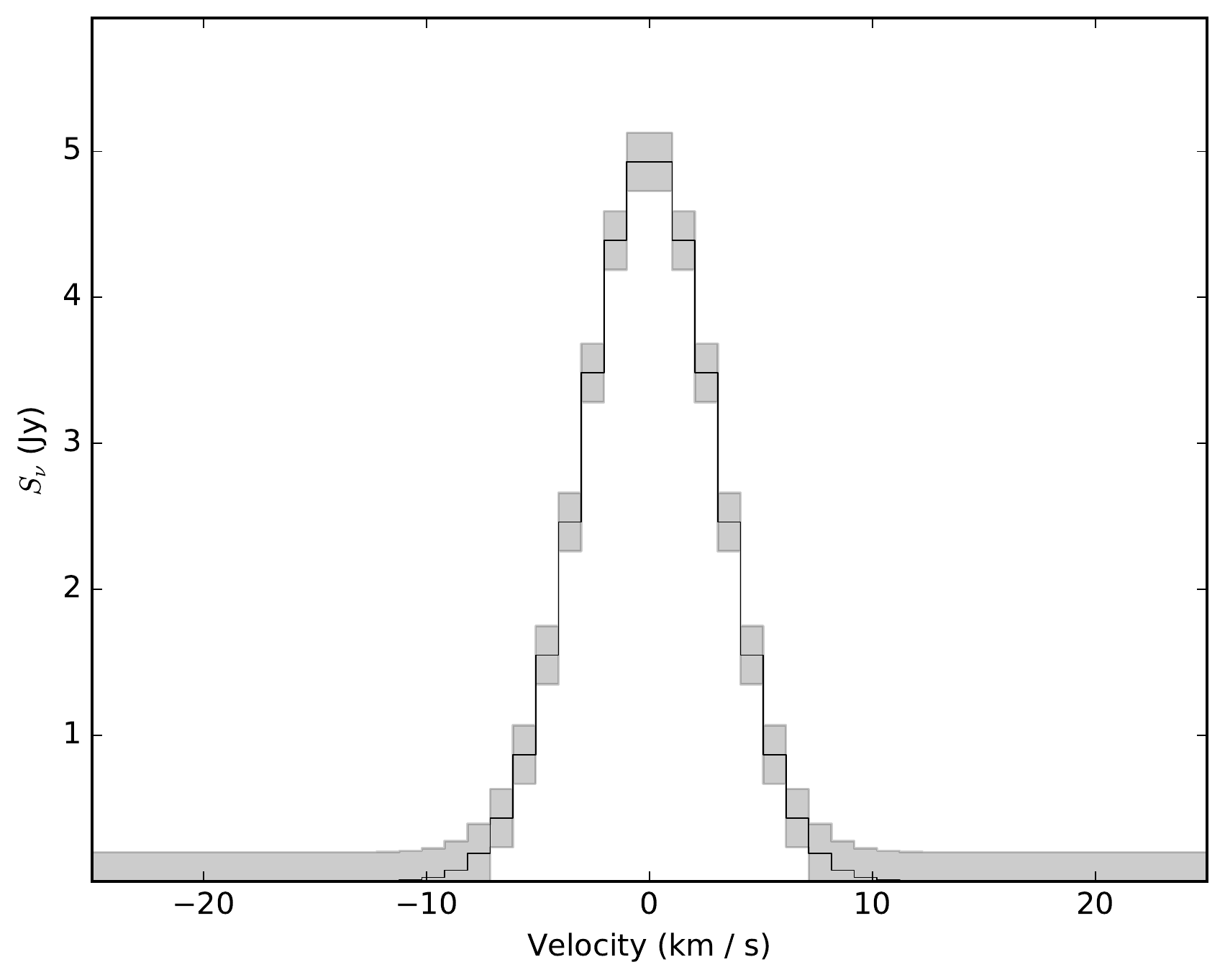}
\caption{An example plotted spectrum showing the automated unit labeling
and errors.  The errors are shown with the \texttt{`fill'} style
and represent symmetric $1-\sigma$ Gaussian errors.  }
\label{fig:example1}
\end{figure}

\subsubsection{Automatic Error Estimation}
\label{sec:autoerror}
In the case where there are portions of the spectrum that have no
signal peaks, a common approach in spectroscopy is to estimate the
errors from the standard deviation of those signal-free pixels.  This approach
assumes the noise is constant across the spectrum.

If all of the peaks present
in the spectrum are fitted well by the model,
the standard deviation of the residual spectrum from the model fit will
accurately represent the uniform errors.     If a fit is performed with
uninitialized errors, they will initially default to a constant value of unity. 
\pyspeckit will then automatically replace the errors with the
standard deviation of the residuals, so they remain constant but will have a value that is related to the data.
This means that performing a fit on the
same data (without associated errors) twice will result in the same parameter
values both times but different errors the second time.

\subsubsection{Parameter error estimation}
\label{sec:parerrest}
Parameter errors are adopted from the \texttt{mpfit} or \texttt{lmfit} fit
results.  The Levenberg-Marquardt algorithm finds a local minimum in parameter
space, and one of its returns is the parameter covariance matrix.  This
covariance matrix is not directly the covariance of the parameters, and must be
rescaled to deliver an approximate error.

The standard rescaling is to multiply the covariance by the sum of the squared
errors divided by the degree of freedom of the fit, usually referred to
as $\chi^2/N$.  The number of degrees of freedom is assumed to be equal to the
number of free parameters, e.g., for a one-dimensional Gaussian, there would be
three: the amplitude, width, and center.  This approach implicitly assumes that
the
model describes the data well and is an optimal fit.  It also assumes that
the model is linear with all of the parameters, at least in the region immediately
surrounding the optimal fit.  These requirements are frequently not satisfied;
see \citet{Andrae2010a} and \citet{Andrae2010b} for details.
We show a demonstration of this approximation process in Appendix \ref{appendix:parerrest}
for the case of a simple Gaussian line profile.


\section{Graphical Design}
\label{sec:gui}
\subsection{GUI development}
Many astronomers are familiar with IRAF's \texttt{splot} tool, which is useful
for fitting Gaussian profiles to spectral lines.  It uses keyboard interaction
to specify the fitting region and guesses for fitting the line profile, but for
most use cases, these parameters could \emph{only} be accessed through the GUI.

The fitting GUI in \texttt{pyspeckit} was built to match \texttt{splot}'s
functionality but with additional means of interacting with the fitter.  In
\texttt{splot}, reproducing any given fit is challenging, since subtle changes
in the cursor position (i.e., the input guess) can significantly change the fit
result.  In \pyspeckit, it is possible to record the results of fits
programmatically and re-fit using those same results.

The GUI was built using \texttt{matplotlib}'s canvas interaction tools.  These
tools are limited to the GUI capabilities that are compatible with all platforms
(e.g., Qt, Tk, Gtk) and therefore exclude some of the more sophisticated fitting
tools found in other software \citep[e.g., \texttt{glue};][]{Beaumont2014b}. 

An example walking through typical interactive GUI usage is in the online
documentation at \url{https://pyspeckit.readthedocs.io/en/latest/interactive.html}.

\subsection{Plotting}
Plotting in \pyspeckit is designed to provide a short path to
publication-quality figures.  The default plotting mode uses histogram-style
line plots and labels axes with \LaTeX-formatted versions of units.

When the plotter is active and a model is fit, the model parameters are
displayed with \LaTeX~formatting in the plot legend.  The errors on the
parameters, if available, are also shown, and these uncertainties are used to
decide on the number of significant figures to display.


\section{Models}
\label{sec:models}
Some of \pyspeckit's internal functions may be replaced by the \astropy\
\texttt{specutils} package in the future.  However, the rich suite modeling in
\pyspeckit is likely to remain useful indefinitely.  This model library
includes some of the most useful general spectral model functions (e.g.,
Gaussian, Lorentzian, and Voigt profiles) and a wide range of specific model
types (e.g., ammonia and formaldehyde hyperfine models, the \hh rotational
ladder, and recombination line models).  These models can be easily used
within \pyspeckit, but they can also be used completely independent from
it (e.g., \url{https://nestfit.readthedocs.io/en/latest/quickstart.html}, \citealt{brian_svoboda_2021_4470028}).
Several of the models rely on \texttt{scipy} \citep{Virtanen2020} for either special functions or multidimensional interpolation.

The base \texttt{Model} class and fitting framework in \pyspeckit provide some
generally useful features that do not need to be re-implemented.  Any model is
generalized to a multi-component form automatically; e.g., the Gaussian model
only describes a single Gaussian spectral component, but the fitting tools
allow any number of independent Gaussians to be fit.

Models are customizable, and examples of registering a new or modified
model in \pyspeckit are included in the online documentation.

A list of the included models, and their parent class when relevant,
is given in Table \ref{tab:models}

\begin{deluxetable}{ccc}
\tablecaption{Summary of Models\label{tab:models}}
\tablewidth{0pt}
\tablehead{
\colhead{Model} & \colhead{Module Name} & \colhead{Parent class} 
}
\startdata
    Hyperfine        & \texttt{hyperfine}              & \nodata\\
    Gaussian         & \texttt{inherited\_gaussfitter} & \nodata\\
    Lorentzian       & \texttt{inherited\_lorentzian}  & \nodata\\
    Voigt            & \texttt{inherited\_voigtfitter} & \nodata\\
    NH$_3$           & \texttt{ammonia}                & Hyperfine  \\
    NH$_2$D          & \texttt{ammonia}                & Hyperfine  \\
    N$_2$H$^+$       & \texttt{n2hp}                   & Hyperfine  \\
    N$_2$D$^+$       & \texttt{n2dp}                   & Hyperfine  \\
    DCO$^+$          & \texttt{dcop}                   & Hyperfine  \\
    LTE Molecule     & \texttt{lte\_molecule}          & \nodata\\
    H$_2$CO (cm)     & \texttt{formaldehyde}           & Hyperfine  \\
    H$_2$CO (mm)     & \texttt{formaldehyde\_mm}       & \nodata\\
    Hydrogen         & \texttt{hydrogen     }          & \nodata\\
    \enddata
\end{deluxetable}


\paragraph{Hyperfine Line Models}
In radio and millimeter spectroscopy, there are many molecular line groups
that are well modelled as Gaussian profiles separated by fixed frequency
offsets.  These hyperfine line groups are often unique probes of physical
parameters because these features have different, known relative optical
depths.  In this case, the measured relative amplitudes of these different
features allow the optical depth (and, in turn, the column density) to be
measured from a single spectrum.  \pyspeckit provides the \texttt{hyperfine}
model class to handle this class of molecular line transitions, and it includes
several molecular species implementations (e.g., HCN, N$_2$H$^+$, NH$_3$, H$_2$CO). 
In this implementation, the excitation temperature of each of the hyperfine components is assumed to be the same, which is the most commonly used assumption but may be violated in some cases.

\section{Cubes}
\label{sec:cubes}
Spectral cubes have become important in radio astronomy, since they are the
natural data products produced by interferometers like ALMA and the JVLA.
Optical and infrared data cubes are growing
more common from integral field units (IFUs) like MUSE on
the VLT, OSIRIS on Keck, NIFS on Gemini, and NIRSpec and MIRI on JWST.

The cube visualization tools built into pyspeckit are limited to spectral and spatial plots.
The \texttt{mapplot} viewer makes a 2D image of a slice of the cube, if given a slice number,
or a projection along the spectral axis if given a function or function name.
Once active, the mapplot viewer can be used interactively: clicking on a pixel
will display that pixel's spectrum in a separate window; clicking and dragging
will produce a circular region whose average spectrum will be plotted in that
window.  The plots shown in the separate window correspond to a spectrum accessible
as the cube's \texttt{.spectrum} object. This basic interaction allows for data exploration,
but is not efficient for fitting each spectrum of a cube.

While many cube operations are handled well by \texttt{numpy}-based packages
like \texttt{spectral-cube} \citep{Robitaille2016,Ginsburg2019}\footnote{\url{https://spectral-cube.readthedocs.io}},
it is sometimes desirable to fit a profile to each spectrum
in a cube.  The \texttt{Cube.fiteach} method is a tool for automated line
fitting that includes parallelization of the fit.  Examples can be
found in the online documentation. 
This tool  has  seen significant use in 
custom made survey pipelines, where the library of spectral models is particularly useful
\citep[e.g,][\url{https://github.com/GBTAmmoniaSurvey/GAS}]{Friesen2017}. 
 It has also been incorporated into other tools, e.g.,
\texttt{multicube}\footnote{\url{https://github.com/vlas-sokolov/multicube}}, SCOUSE \citep[][see Appendix \ref{appendix:scouse}]{Henshaw2016,Henshaw2019}, \texttt{make\_cube} \citep{Youngblood2016}, and \texttt{pyspecnest}\footnote{\url{https://github.com/vlas-sokolov/pyspecnest}} \citep{Sokolov2020-pyspecnest}.


\section{Summary}
\label{sec:summary}

\texttt{pyspeckit} is a versatile tool for spectroscopic analysis in python and
is one of the \astropy affiliated packages.
%


\acknowledgments
JEP acknowledges the support by the Max Planck Society. 
AG acknowledges support from the NSF under grant AST 2008101.

\input{solobib}

\appendix
\section{Parameter Error Estimation for a simple 1D Gaussian profile}
\label{appendix:parerrest}
As discussed in Section \ref{sec:parerrest}, parameter errors are estimated in
\pyspeckit by the underlying \texttt{lmfit} or \texttt{mpfit} tools using the
approximation that the reduced chi-squared is unity, $\chi^2/n=1$.  We
demonstrate here that, for a simple one-dimensional Gaussian profile, this
approximation results in an excellent recovery of the
underlying parameter errors.

\begin{figure*}[!htp]
\includegraphics[scale=1,width=7in]{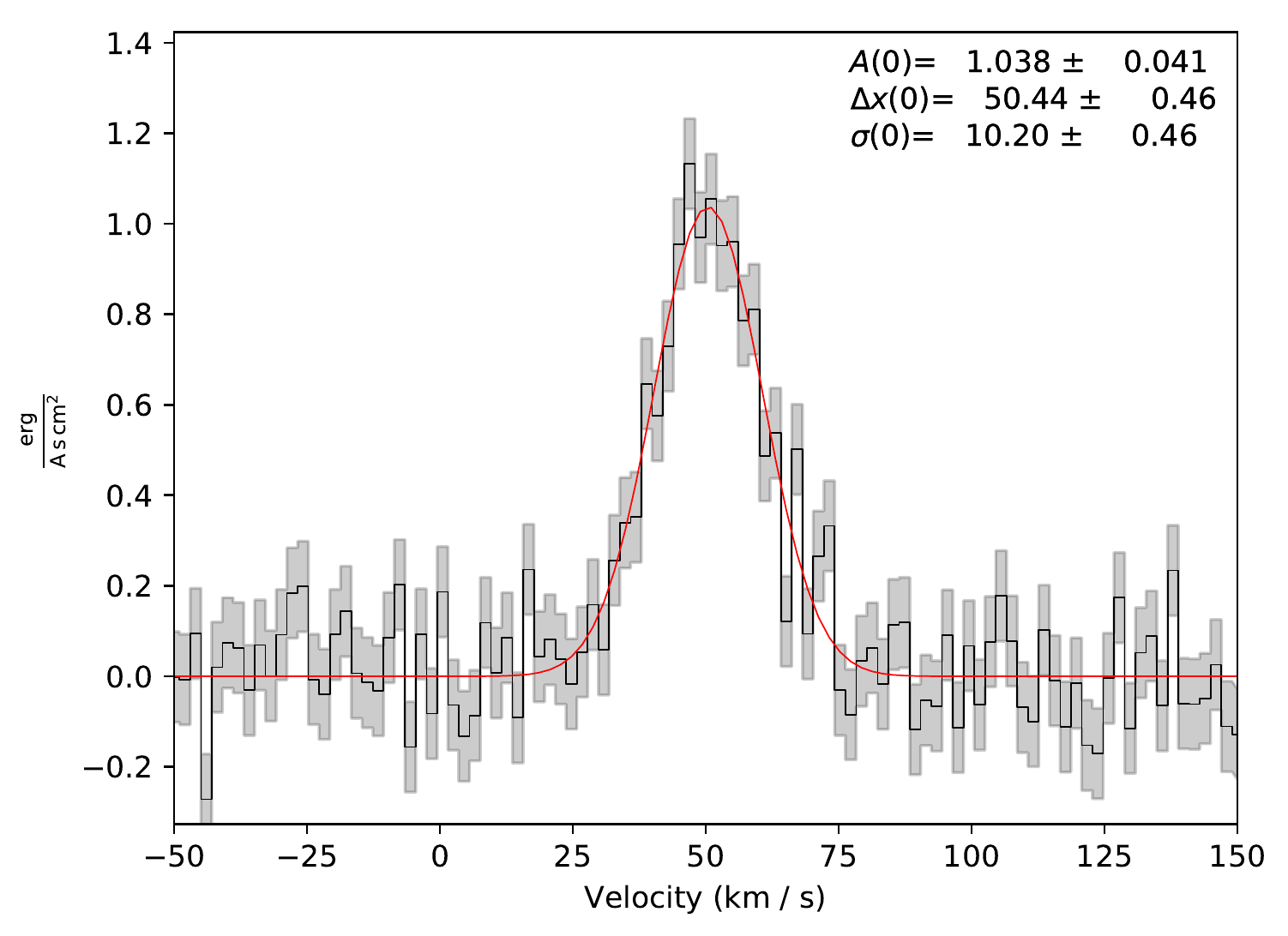}
\caption{One-dimensional Gaussian profile fit to a synthetic spectrum.
The parameter values and errors are shown in the upper right.  The number of
significant figures displayed in both the value and the error is automatically
set to one digit more than the last significant digit in the error.}
\label{fig:synthspecdemo}
\end{figure*}

In Figure \ref{fig:synthspecdemo}, we show a synthetic spectrum with uniform
Gaussian random noise and perfectly-measured uncorrelated data errors.
The fitted model is a one-dimensional Gaussian profile with free parameters
amplitude, center, and width.  The fit results are given in the figure.

To produce a good error estimate under the $\chi^2/n=1$ approximation, the error
distribution must be Gaussian, the model must be
linear in all parameters, and the model must be the correct underlying model
\citep{Andrae2010b}.

\begin{figure*}[!htp]
\includegraphics[scale=1,width=7in]{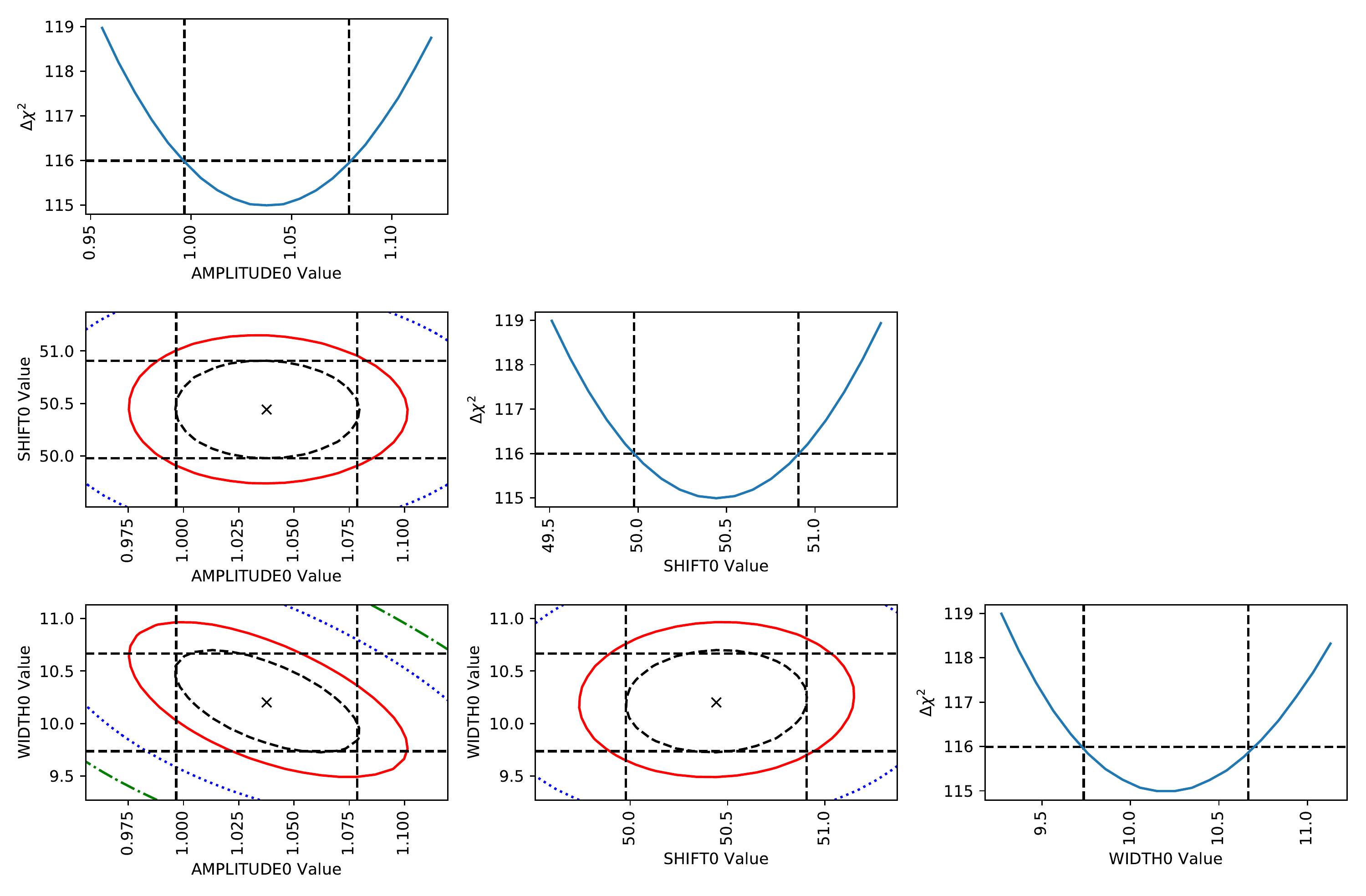}
\caption{Error estimate figure.
In all panels, the vertical
dashed lines show the estimated $1\sigma$ errors from the optimizer, while the
horizontal dashed lines show the value $\Delta\chi^2=1$, which corresponds to
the 68\% confidence interval for that parameter.
In the off-diagonal panels, contours are shown at $\Delta\chi^2=1,2.3,6.2,11.8$ (black dashed, red, blue dotted, green dash-dot),
corresponding to 39.3\%, 68\%, 95\%, and 99.5\% confidence
regions ($1\sigma$ for a single parameter,
then $1\sigma$, $2\sigma$, and $3\sigma$ for two parameters).
See Appendix \ref{appendix:parerrest} for details and interpretation.}
\label{fig:parerrestdemo}
\end{figure*}

Figure \ref{fig:parerrestdemo} shows the $\chi^2$ values in parameter space
surrounding the best-fit value.  Along the diagonal, we show the $\chi^2$
values for the individual parameters with all others marginalized over
by taking the minimum $\chi^2$ value over the explored parameter space.  The vertical
dashed lines show the estimated $1\sigma$ errors reported by the \texttt{mpfit}
optimizer, while the
horizontal dashed lines show the value $\Delta\chi^2=1$, which corresponds to
the 68\% confidence interval for that parameter.  If the $\chi^2/n=1$
approximation is perfect, the dashed lines should intersect with the blue curves, and in this case, they do for all parameters.  

Off of the diagonal of Figure \ref{fig:parerrestdemo}, we show the
two-dimensional marginal distributions.  Contours
are shown at
$\Delta\chi^2=1,2.3,6.2,11.8$, corresponding to 39.3\%, 68\%, 95\%, and 99.5\%
confidence regions
(the first value corresponds to $1\sigma$ uncertainty when marginalized to a single parameter, while the others correspond to $1\sigma$, $2\sigma$, and
$3\sigma$ for two parameters, respectively, assuming a normal
distribution) .  The
vertical and horizontal dashed lines show the estimates from the $\chi^2/n=1$
approximation for a \emph{single} parameter; these are  expected to intersect
the innermost 1$\sigma$ contour  when marginalized to a single parameter. 
The shift vs amplitude and shift vs width diagrams are both well-behaved, with error
contours tracing out a symmetric distribution centered on the true parameters marked
with an `x'.
However, the width vs amplitude plot indicates that the single-parameter
errorbars 
are partly driven by the significantly correlated uncertainty between 
these parameters.
This information is captured in the covariance
matrix that is used to compute the single-parameter errors, as it has
significant values in the off-diagonal parts of the matrix.  

The source code for this example can be found in the \pyspeckit github
repository in \texttt{examples/synthetic\_spectrum\_example\_witherrorestimates.py}\footnote{\url{https://github.com/pyspeckit/pyspeckit/blob/master/examples/synthetic_spectrum_example_witherrorestimates.py}}.

\section{Parameter Error Estimation for Ammonia}
\label{appendix:parerrestammonia}
In Appendix \ref{appendix:parerrest}, we showed the parameter estimation results
in the case of a modeled 1-dimensional Gaussian.  One of the most commonly used
models in \pyspeckit is the ammonia (\ammonia) hyperfine model, which has several
additional emission lines and several parameters governing those lines.

The ammonia inversion transitions are notable for having spectrally resolved
hyperfine components under typical Galactic molecular gas conditions, the relative weights of which are governed by quantum
mechanics \citep{Mangum2015}. The existence of these additional components often
allows for direct estimates of the optical depth of the central line, which is
optically thicker than the other components, thereby making column density
estimates from a single spectral band relatively straightforward.

The model for these lines is more complicated than that for a single Gaussian.
The model must include a simplified version of the radiative transfer equation
and must simultaneously produce the predicted emission of several lines.
Additionally, there are several approximations for the relative line strengths that are convenient to use
under different circumstances, so \pyspeckit implements several different
variants of the \ammonia model.

In this section, we show parameter estimates analogous to those in Section
\ref{appendix:parerrest}.
We examine a case where the fitted lines are in local
thermodynamic equilibrium (LTE), such that the ratios of the $(J,K)=(1,1)$
to $(2,2)$ line is governed by the rotational temperature $T_\mathrm{R}$ but
the individual lines both have $T_{\mathrm{ex}}=T_{\mathrm{R}}$.

The free parameters in the ammonia model are the rotational temperature,
$T_{\mathrm{R}}$, which governs the relative populations of the rotational
states, the excitation temperature $T_{\mathrm{ex}}$, which governs the
relative populations of the two levels within a single inversion transition,
the column density, $N(\ammonia)$, which specifies the total column density of
\ammonia integrated over all states (note that this parameter enters the model
as $10^N$, i.e., we optimize the log of the column density), the line-of-sight
velocity $v_\mathrm{LoS}$, the line width $\sigma_v$, and the ortho-to-para
ratio parameterized as the fraction of ortho-\ammonia $F_{ortho}$.  In the
examples below, we fix $F_{ortho}=0$ and treat only para-\ammonia lines.

\begin{figure*}[!htp]
\includegraphics[scale=1,width=7in]{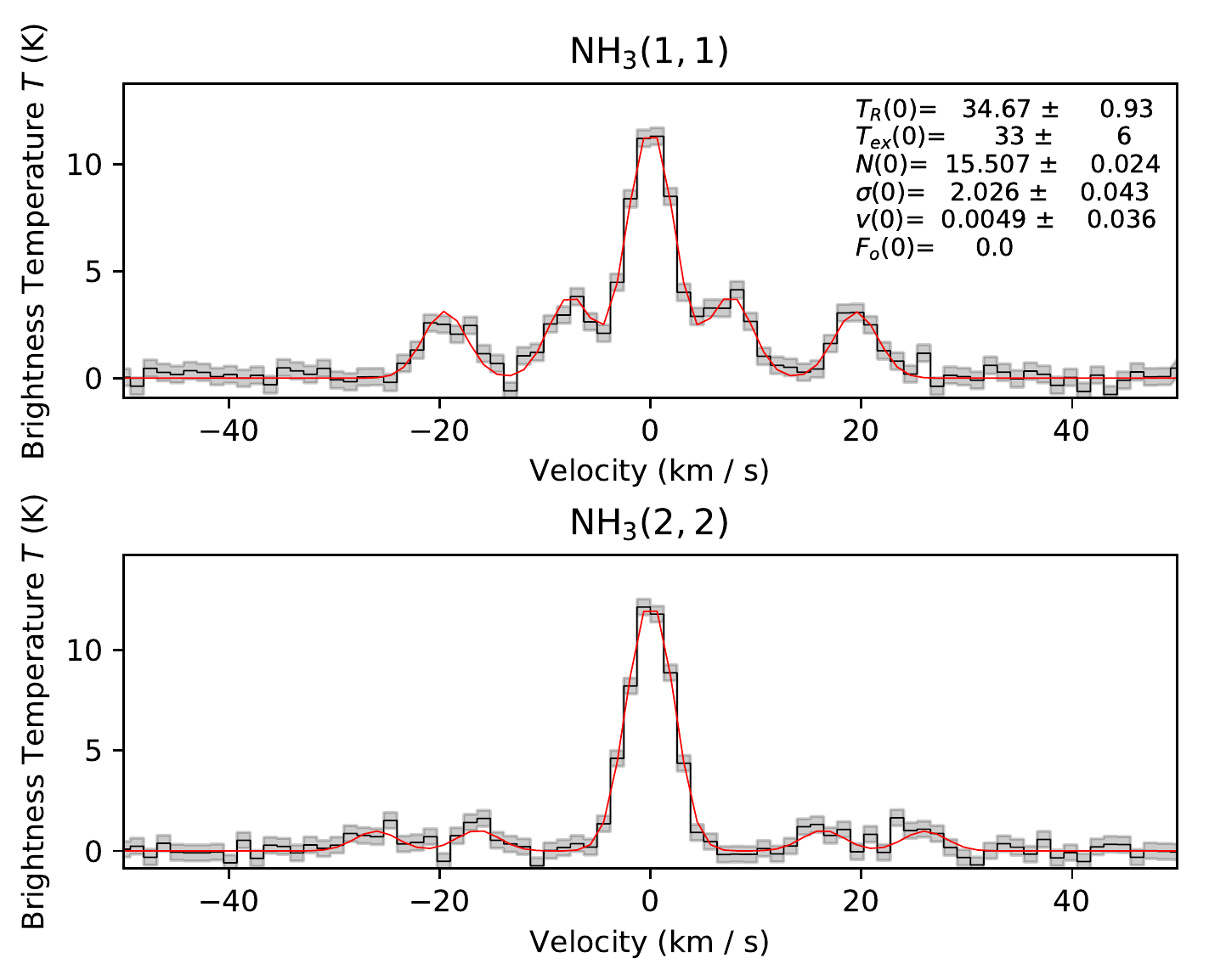}
\caption{Ammonia model profile fit to a synthetic spectrum.
The parameter values and errors are shown in the upper right.  The associated
error estimate triangle diagram is shown in Figure \ref{fig:nh3parerrestdemo}.
The correct parameters are $T_{\mathrm{R}}=T_{\mathrm{ex}}=35$, $N=15$, $\sigma_v=2$, and $v=0$,
all of which are reasonably recovered.  However, note that $T_{\mathrm{ex}} > T_{\mathrm{R}}$
is generally nonphysical, yet the allowed parameter space for $T_{\mathrm{ex}}$ includes
such values. 
This two-panel plot was produced automatically using the
\texttt{pyspeckit.wrappers.fitnh3.plot\_nh3} command.
}
\label{fig:nh3synthspecdemo}
\end{figure*}

\begin{figure*}[!htp]
\includegraphics[scale=1,width=7in]{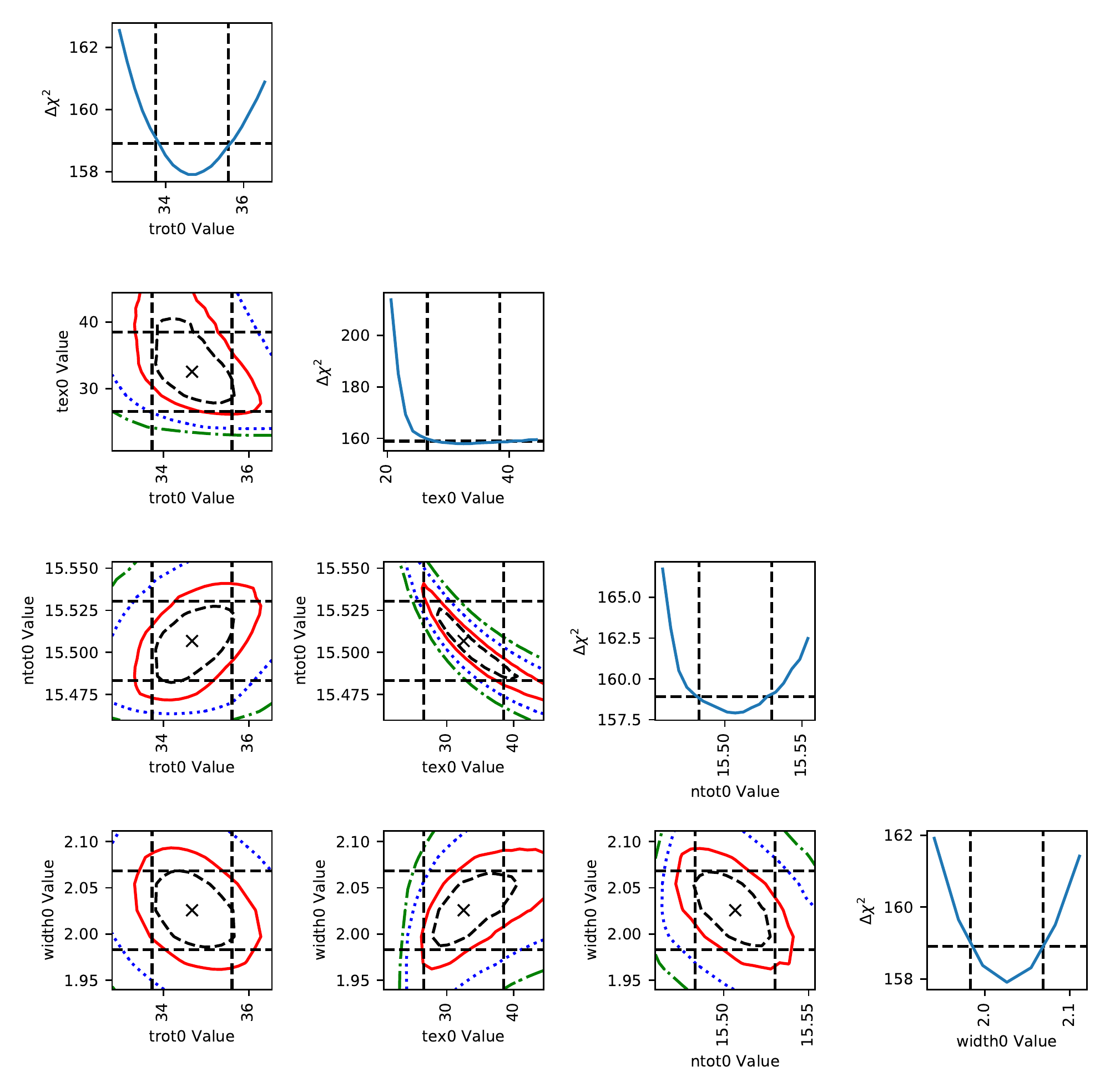}
\caption{Error estimate figure for the default \ammonia model.  The panels
are labeled as in Figure \ref{fig:parerrestdemo}.  The centroid velocity panel
is excluded from this plot because it is independent of the other parameters,
exhibiting no degeneracy.  The most
relevant panel is the ntot0 vs tex0 panel, which plots the column $N$ against
the excitation temperature $T_{\mathrm{ex}}$: both of these parameters govern the peak
amplitude of the spectrum, so they are degenerate.  
The vertical and horizontal dashed lines represent the $\delta\chi^2=1$ and $\pm1\sigma$
computed errors, respectively.  Their intersection points also intersect
the blue marginalized $\delta\chi^2$ curve in all of the displayed cases, indicating
that the approximation of the errors using the $\chi^2/n=1$ assumption is excellent.
However, the asymmetry seen in the blue $T_{ex}$ curve suggests that the approximation
for that parameter is likely to break down in some cases because the true
errors are asymmetric.
}
\label{fig:nh3parerrestdemo}
\end{figure*}

The fit results from the first case are shown in Figures
\ref{fig:nh3synthspecdemo} and \ref{fig:nh3parerrestdemo}.  The fit recovers
the input parameters, but reveals one of the important caveats when using any
optimization algorithm: in some models, parameters are degenerate, and
therefore using the diagonal of the covariance matrix to estimate the variance
can result in incorrect error estimates.  While the errors on most parameters
appear reasonable, there is a very large error on the excitation temperature
$T_{\mathrm{ex}}$, which is driven by the degeneracy of $T_{\mathrm{ex}}$ with
$N_\mathrm{tot}$.  The asymmetry of the error on $T_{\mathrm{ex}}$ is apparent
in Figure \ref{fig:nh3parerrestdemo}, but it is not captured by the optimizer's
reported error results; the asymmetry occurs because $T_{ex}$ is in the
exponent in the model equations.

In such situations, it can be beneficial to measure the parameter errors
in different ways.  Using the \texttt{emcee} and \texttt{pymc} wrappers
can help do this.  Examples of how to use these Monte Carlo samplers
to acquire better parameter errors once an optimization has already
been performed are available in the online documentation:
see \url{http://pyspeckit.readthedocs.io/en/latest/example_pymc.html}.

More sophisticated examples, including fitting of a non-LTE ammonia spectrum
in which $T_{\mathrm{ex}} < T_{R}$, are available in the example directory
of \pyspeckit (\url{https://github.com/pyspeckit/pyspeckit/tree/master/examples}),
specifically
\url{https://github.com/pyspeckit/pyspeckit/tree/master/examples/synthetic_LTE_ammonia_spectrum_example_witherrorestimates.py}
and
\url{https://github.com/pyspeckit/pyspeckit/tree/master/examples/synthetic_nLTE_ammonia_spectrum_example_witherrorestimates.py}.

These examples also include demonstrations of how to force the optimizer to
ignore nonphysical values while still obtaining useful constraints on the free
parameters.  Constrained fitting approaches can be helpful in cases like the
ammonia fit, where high values of $T_{ex}$ where $T_{ex}>T_{rot}$ are
statistically likely given the model, but physically disallowed; constrained
fitting allows the known physical limits to rule out bad portions of parameter
space.

\section{Comparison of N$_2$H$^+$ (1-0) results with CLASS}
\label{appendix:N2Hp}

One of the most frequently used line transitions for the study of dense gas
kinematics is N$_2$H$^+$ (1-0) at 93.17\,GHz. 
The transition displays several hyperfine components with well determined 
relative frequencies and weights. 
The standard approach for analyzing this line has been to use the \verb+HFS+
mode within CLASS. 
Here we show that using the N$_2$H$^+$ \texttt{hyperfine} model in \pyspeckit,
we obtain the same results in both optically thin and thick models.

The main difference between the CLASS and \pyspeckit parametrization is that
the former does not report excitation temperature ($T_{ex}$), but the area of the
line profile. The reported area is $\tau \times T_{ant}$, where
\begin{equation}
T_{ant}=J(T_{ex}) - J(T_{background})~,
\end{equation}
where
\begin{equation}
J_{\nu}(T) = \frac{h\,\nu}{k_B} \frac{1}{(e^{h\,\nu/k_B\,T} -1)}~.
\end{equation}
We derive the equivalent $T_{ex}$ from the reported line fit parameters. 
Moreover, in the optically thin case both fits are performed using the common
assumption of $\tau=0.1$ as a fixed parameter.

Tables \ref{tab:n2hpthin} and \ref{tab:n2hpthick} show the results of fitting
an example spectrum in both CLASS and \pyspeckit.  The resulting fits differ by
$<1\%$ in most parameters, with a slightly greater discrepancy in the velocity
centroid but consistent within the reported fit uncertainties.


\begin{deluxetable}{cccc}
    \label{tab:n2hpthin}
\tablecolumns{4}
\tablecaption{Best fit parameters in optically thin model (3 parameters)\label{table-thin}}
\tablehead{\colhead{Parameter} &\colhead{Input value} & \colhead{pyspeckit fit} & \colhead{CLASS fit}
}
\startdata
T$_{ex}$      & 9.0   & 3.454$\pm$0.014     & 3.451\\
$V_c$          & 0.0   & 0.0016              & 0.0028$\pm$0.0068\\
$\sigma_v$  & 0.3   & 0.2942$\pm$0.0062 & 0.2930$\pm$0.0060\\
$\tau_{all}$  & 0.01  & 0.1 & 0.1 \\
$Area$ & & & 0.0607$\pm$0.0012\\
\enddata
\end{deluxetable}

\begin{deluxetable}{cccc}
    \label{tab:n2hpthick}
\tablecolumns{4}
\tablecaption{Best fit parameters in optically thick model (4 parameters)\label{table-thick}}
\tablehead{\colhead{Parameter} &\colhead{Input value} & \colhead{pyspeckit fit} & \colhead{CLASS fit}
}
\startdata
T$_{ex}$     & 9.0  & 9.19$\pm$0.19     & 9.1833\\
$V_c$         & 0.0  & 0.00042 & -0.000$\pm$0.0063\\
$\sigma_v$ & 0.3  & 0.3047$\pm$0.0062 & 0.3041$\pm$0.0063\\
$\tau_{all}$ & 9.0  & 8.13$\pm$0.59   & 8.10$\pm$0.59\\
$Area$ & & & 49.0$\pm$2.25\\
\enddata
\end{deluxetable}

\section{LTE Model}
\label{appendix:ltemodel}

\pyspeckit includes tools to model generic local thermodynamic equilibrium (LTE) emission lines.
The modeling tools include wrappers to retrieve the appropriate molecular line parameters from the JPL \citep{Pickett1998} or CDMS\footnote{\url{https://spec.jpl.nasa.gov/}} \citep{Muller2005} spectroscopy databases via their astroquery interfaces \citep{Ginsburg2019}.

The main modeling function, \texttt{LTE\_molecule.generate\_model}, operates on the combination of line parameters (centroid, width), physical parameters (excitation temperature $T_{ex}$, total column density $N_{tot}$), and molecular line parameters (rest frequency $\nu_{rest}$, Einstein A value $A_{ij}$, degeneracy $g_u$, upper state energy level $E_U$, and the partition function $Q(T_{ex})$).
It implements the equations:
\begin{equation}
            \tau_\nu = \frac{c^2}{8 \pi \nu^2} A_{ij} N_u \phi_\nu
                    \left[ \exp\left( \frac{h \nu}{k_B T_{ex}} \right)  - 1 \right]
\end{equation}
where $c$ is the speed of light, $k_B$ is Boltzmann's constant, $h$ is Planck's constant, $N_u$ is given by
\begin{equation}
        N_{u} = N_{tot} \frac{g_u}{Q} \exp\left(\frac{-E_u}{k_B T_{ex}} \right)
\end{equation}
and $\phi_\nu$ is assumed to be a Gaussian such that
\begin{equation}
        \phi_\nu = \frac{1}{\sqrt{2 \pi} \sigma_\nu}
        \exp \left[ -\frac{(\nu-\nu_0)^2}{2 \sigma_\nu^2} \right]
\end{equation}
where $\sigma_\nu$ is the Gaussian line width (not the FWHM) in frequency units, $\sigma_\nu = \frac{\sigma_v}{c} \nu_{rest}$ and $\sigma_v$ is the Gaussian line width in velocity units.
The above is based on Equations 11 and 29 of \citet{Mangum2015}.
The returned spectral model is then calculated as 
\begin{equation}
    \begin{array}{lll}
    I_\nu & = & J_\nu(T_{ex}) \left[1 - e^{-\tau}\right] + J_\nu(T_{BG}) e^{-\tau}  - J_\nu(T_{BG})  \\
          & = & \left[J_\nu(T_{ex})-J_\nu(T_{BG})\right] \left[1 - e^{-\tau}\right] \\
    \end{array}
\end{equation}
where $T_{BG}$ is the background temperature (assumed 2.73 for the default CMB temperature).
We have split the equation into two lines to indicate that the return is explicitly background-subtracted.
$J_\nu$ is the Rayleigh-Jeans equivalent brightness temperature
\begin{equation}
    J_\nu = \frac{h \nu}{k_B} \left[\exp\left(\frac{-h \nu }{k_B T}\right) -1 \right]^{-1}
\end{equation}.
The partition function can be specified as an array, which enables modeling lines from different species simultaneously.

To verify the accuracy of these models, we benchmarked the \pyspeckit models against both XCLASS \citep{Moller2017} and molsim \citep{Lee2021}.
The benchmark notebook is provided here: \url{https://github.com/pyspeckit/pyspeckit-tests/blob/master/xclass-benchmark/CO_Benchmark.ipynb}.
We modeled the lowest 12 transitions of CO.
We obtain good agreement ($<1\%$ difference) with XCLASS at the line peak for all transitions when using option \texttt{Interf\_Flag = False}.
In the low-J transitions, we see significant fractional differences in the line wings, but with very small absolute values (the difference in the integral is $<10^{-4}$); these appear to come from rounding errors in the evaluation of the Gaussian function.
These small differences are very unlikely to affect any modeling.
We obtain significantly different answers with \texttt{Interf\_Flag = True}; in this case, XCLASS returns a peak that is lower by the CMB temperature.

We additionally benchmarked two more complicated molecules, CH$_3$OH and CH$_3$CN, and again found excellent agreement.


\section{\texttt{ScousePy}}
\label{appendix:scouse}

\texttt{ScousePy} is a tool for the decomposition of data cubes with multi-component spectral lines, and makes use of \pyspeckit's fitting functionality (\S \ref{sec:fitters}). Broadly speaking, spectral decomposition algorithms can be divided into two classes: bottom-up and top-down. The former treat the decomposition of individual spectra independently from one another, while the latter approach uses spatial averaging to estimate initial guesses for the decomposition of individual spectra. \texttt{ScousePy} follows the latter of these two approaches.  \texttt{ScousePy}
\begin{enumerate}
    \item 
breaks a data cube into sub-regions of user-defined size.  A spatially averaged spectrum is extracted from each sub-region. 
    \item uses  derivative spectroscopy to estimate the number of components and the properties of those individual components in each spatially-averaged spectrum. These initial guesses are passed to \pyspeckit's \texttt{specfit} to perform the fit. 
    \item uses the output solution (\texttt{specfit.fitter.params}) from the average spectrum as new initial guesses for fitting the individual spectra contained within the averaging region.
    \item displays the results in an interface that enables quality assessment and re-fitting if necessary.
\end{enumerate}

\texttt{ScousePy} makes use of \pyspeckit's \texttt{specfit} in two main ways.
First, non-interactively, with initial guesses supplied to \texttt{specfit} via \texttt{specfit.guesses}.
Second, \texttt{ScousePy} also makes use of \pyspeckit's interactive fitting functionality, which uses keyboard and mouse interaction to provide \texttt{specfit.guesses}.

\begin{figure*}[!htp]
\includegraphics[scale=1,width=7in]{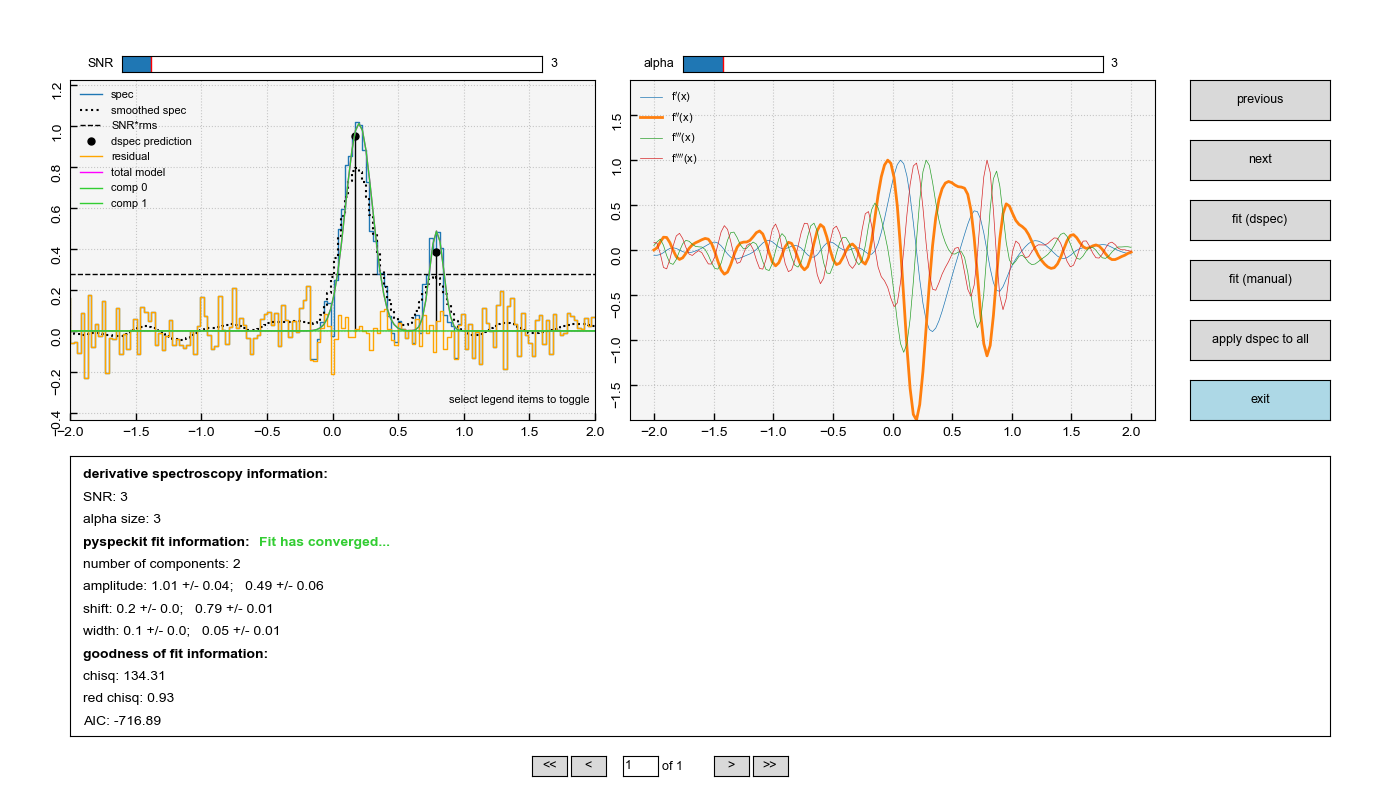}
\caption{Gaussian decomposition of a two-component spectrum performed by \texttt{ScousePy} using \pyspeckit. The top-left panel shows the spectrum (black), the same spectrum smoothed with a Gaussian 1D kernel (dotted black line) using \texttt{astropy.convolution.Gaussian1DKernel}, which is used for the derivative spectroscopy (right panel), the location and amplitude of the initial guesses determined from derivative spectroscopy (black lollipops) and supplied to \texttt{specfit}, the best fitting model solution (green and magenta lines indicate individual components and the total model), the residual spectrum after subtracting the model (orange line), and the $3\sigma$ noise level determined from \texttt{ScousePy.getnoise}. The top-right panel shows the first to fourth derivatives of the smoothed spectrum (dotted black line).
These derivatives are used to estimate the number of components and their amplitudes, widths, and centroids (\texttt{ScousePy.DSpec}). The information panel at the bottom shows the output information from \pyspeckit. This information includes the model parameters and their uncertainties and goodness-of-fit statistics. The buttons on the right can be used to control \pyspeckit \ either non-interactively using ``fit (dspec)'' or interactively using ``fit (manual)''. }
\label{fig:scousepydemo}
\end{figure*}

An example of the \texttt{ScousePy.ScouseFitter} GUI is shown in Figure \ref{fig:scousepydemo}. This figure shows the Gaussian decomposition of a two component spectrum. \texttt{ScousePy} provides initial guesses to \pyspeckit's \texttt{specfit} using derivative spectroscopy, in which:
\begin{enumerate}  
\item the input spectrum is smoothed using \texttt{astropy.convolution.Gaussian1DKernel} (see \texttt{ScousePy.DSpec}).
\item the first to fourth order derivatives of the smoothed spectrum are computed and used to determine the number of components and their amplitudes, centroids, and widths (see lollipops in the top left panel). 
\end{enumerate}
These values are then supplied to \texttt{specfit} via \texttt{specfit.guesses}. The fit is performed using \texttt{lmfit}, and the result is displayed in the information panel. The buttons on the right hand side of the GUI can be used to enter \pyspeckit's interactive fitter functionality via ``fit (manual)'' \footnote{Note that this example analysis has been performed using \texttt{ScousePy.ScouseFitter} in its stand-alone mode, which can be used for the decomposition of an individual spectrum or lists of spectra. The same procedure is followed during cube fitting.}. The relevant fit information produced by \pyspeckit (or further derived from the quantities output by \texttt{specfit}) are located in a dictionary called \texttt{ScousePy.ScouseFitter.modelstore}. Further information on the \pyspeckit \ methodology used by \texttt{ScousePy} can be found in the documentation for \texttt{ScousePy.ScouseFitter} and \texttt{ScousePy.Decomposer} (see \url{https://scousepy.readthedocs.io/en/latest/}). 

\end{document}

%% file: solobib.tex
\bibliographystyle{aasjournal}
\bibliography{extracted}